\documentclass[runningheads]{llncs}

\usepackage[T1]{fontenc}
\usepackage{amsmath}
\usepackage{graphicx}
\usepackage{seqsplit}
\usepackage{tikz}
\usetikzlibrary{arrows.meta,positioning,fit,backgrounds}
\usepackage[hidelinks]{hyperref}

\usepackage{etoolbox}
\AtBeginEnvironment{thebibliography}{\setlength{\itemsep}{0pt plus 0.2pt minus 0.3pt}}

\begin{document}

\title{Threshold Choice, Not Sample Size, Bounds Trustless Verification of Nondeterministic Compound AI Workflows}
\titlerunning{Verifying Semantic Reproduction of Compound AI on the 3D Continuum}

\author{Alper Alimo\u{g}lu\inst{1}}
\authorrunning{A. Alimo\u{g}lu}
\institute{Independent Researcher\\
\email{alper.alimoglu@gmail.com}}

\maketitle

\begin{abstract}
Compound AI pipelines chain LLM calls, retrievers, and tools and are
nondeterministic: sampling, model updates, and volatile tool responses make
one input yield different outputs across runs. Such pipelines increasingly run
across edge, cloud, and orbital nodes owned by no single party, whose
optimizations discard intermediate results before inspection. Verifying
reproduction there means tolerating nondeterministic outputs, a node that may
not report honestly, and intermittent access to any shared record; existing
work addresses at most two at once. We give a protocol covering all three: it
commits digests of each stage's inputs, outputs, and context under a policy
digest pinning the metric and threshold, anchors them without trusting the
executing node, defers under partition, and decides a challenge on the median
of $k$ re-executions, with no quorum. Where that procedure breaks is the main
result. On a synthetic HotpotQA pipeline a calibrated fixed threshold accepts
44 of 45 honest reproductions and rejects 104 of 105 divergent pairs, yet lets
same-input fabrication through in 27 of 29 trials at $k=5$, more samples being
no help since sampling sharpens an estimate without moving it. Holding that
metric and this pipeline's re-execution spread fixed, the binding constraint
is the threshold rather than the sample size: one derived per execution
detects 19 of 29 where the best constant matched to the same zero honest
rejections reaches 9, rejects no honest commitment at $k=5$ though 3 of 15 at
$k=3$, and catches 11 of 15 of an attacker built against it, which has to aim
at a target drawn only after its commitment exists. That rule is measured
rather than deployed.

\end{abstract}

\section{Introduction}

A Compound AI workflow chains LLM calls, retrievers, and external tools into
a multi-stage pipeline~\cite{zhou2026} that does not repeat: the same input
can yield different outputs from sampling, model-version drift, and tool
responses that change over time. Orchestrators such as
HyperDrive~\cite{pusztai2024hyperdrive} place such stages across a 3D
continuum of edge, cloud, and low-earth-orbit nodes owned by no single party,
and Currus~\cite{kaya2025currus} shows the same pressure across edge-cloud
alone. Meeting their latency targets optimizes the data path against
retaining a trace: GoldFish~\cite{marcelino2024goldfish},
Truffle~\cite{marcelino2024truffle}, and
Databelt~\cite{marcelino2025databelt} respectively cache, fold into cold-start
time, and propagate stage state along orbiting nodes, each deleting or
relocating the intermediate results an independent party would need to
confirm reproduction.

The executing node may also not report honestly, and any shared record able to
adjudicate is frequently unreachable. Verification therefore needs tolerance
of legitimate nondeterminism, soundness against a misreporting node, and
correctness despite intermittent connectivity, and prior work supplies at most
two (Section~2). Reproducibility-verification work checks semantic equivalence
rather than bit-exact comparison~\cite{wang2026survey} but trusts the node's
report and assumes a reachable reference. Decentralized-provenance
work~\cite{tunstad2019hyperprov,marchioro2025}, including our own
eBlocBroker~\cite{alimoglu2024eblocbroker_cluster,alimoglu2024eblocbroker_ccpe},
removes the trusted recorder by anchoring
on a ledger but treats the computation as deterministic and the ledger as
always reachable.

This paper makes four contributions. A commitment scheme digests each stage's
inputs, outputs, and context at its boundary, under a policy digest pinning
the metric, threshold, and environment a challenge must use. An anchoring
mechanism binds each commitment to an execution identifier and a per-recorder
predecessor chain; a challenge decides on the median of $k$ re-executions, and
a retention rule makes unserved artifacts \textsc{Unavailable}. A
proof-of-concept on a synthetic HotpotQA pipeline, single-process throughout
(Section~7), measures overhead, threshold accuracy, soundness up to an
adaptive attacker, and partition behaviour, and the main result is where that
soundness boundary lies: in the fixed threshold rather than the sample size, a
per-execution threshold being one an attacker built against it cannot reliably
aim at. Trustless here means no trusted arbiter over the two
properties Section~3.2 claims, commitment integrity and semantic consistency:
a challenge returns a local diagnostic, not an adjudicated verdict, a shared
sample, or an attestation of origin. Both rest on two premises the protocol
assumes rather than enforces (Section~3.4), and what the partition cases
establish is ordered immutability conditional on eventual submission rather
than distributed operation (Section~\ref{sec:partition}).

\section{Related Work}

Three lines of work each address part of this problem
(Table~\ref{tab:relatedwork}). One verifies at the level of meaning.
Zhou detects divergent agentic tool use by semantically comparing a replayed
execution to the original. Wang et al.'s survey scopes out untrusted
recorders, decentralized verification, and intermittent connectivity.
Cankaya~\cite{cankaya2026gpu} pursues bit-exact GPU-inference
verification by controlling hardware nondeterminism.

\begin{table}[htbp]
\caption{No prior line holds all three conditions of Section~1 at once.}
\label{tab:relatedwork}
\centering
\footnotesize
\begin{tabular}{lccc}
\hline
\textbf{Line of work} & \textbf{Nondet.} & \textbf{Recorder} & \textbf{Partition} \\
\hline
Semantic verification~\cite{zhou2026,wang2026survey,cankaya2026gpu} & Yes & No & No \\
Ledger-anchored provenance~\cite{tunstad2019hyperprov,marchioro2025,wittek2021bloxberg,slidechain2024,alimoglu2024eblocbroker_cluster,alimoglu2024eblocbroker_ccpe,autonomoussoftwareorg2026} & No & Yes & No$^{a}$ \\
Continuum placement~\cite{pusztai2024hyperdrive,kaya2025currus} & --- & --- & --- \\
This work & Yes & Yes$^{b}$ & Yes$^{c}$ \\
\hline
\end{tabular}
\vspace{1pt}\\
{\scriptsize $^{a}$Except SwarmDAG~\cite{tran2019swarmdag}.
$^{b}$Semantic consistency only, measured rather than asymptotic; not
execution provenance (Section~3.2).
$^{c}$Ordered immutability conditional on eventual submission, demonstrated in
one process; not distributed retrieval or recovery of lost queued state
(Section~\ref{sec:partition}).}
\end{table}

A second line removes the trusted recorder by anchoring on a ledger.
HyperProv, Marchioro et al., Wittek et al.~\cite{wittek2021bloxberg}, and
SlideChain~\cite{slidechain2024} anchor workflow, data, or semantic
provenance without a central coordinator; the last of these assumes
determinism. Our own eBlocBroker belongs here; it schedules DAG stages via
smart contracts. AutonomousSoftwareOrg~\cite{autonomoussoftwareorg2026}
records executions on-chain, but it hashes them under an assumption of
determinism and names defense against a lying provider as open. A third
line places stages across the continuum without verifying them:
HyperDrive and Currus emit no verifiable record, the optimizations that
make placement fast erasing the trace a verifier needs.

That gap is what this paper addresses. Its soundness holds against divergent
claims, not against fabrication inside the accepted equivalence class.

\section{Problem and Threat Model}

\subsection{System Model and Semantic Reproduction}

A Compound AI workflow is a sequence of stages $s_1, \dots, s_n$: each is
dispatched with an input $x_i$, executes on some node, and produces an output
$y_i$ with an execution context $c_i$, the information beyond $x_i$ the
execution depended on, such as a sampled seed, a model version, or a tool
call's response body. The executing node is neither fixed in advance nor
controlled by the party that later verifies reproduction. Bit-exact comparison of $y_i$ across two executions is not a meaningful
correctness criterion here, since $s_i$ may be nondeterministic by design. We
define reproduction against a stage-appropriate, publicly computable
\emph{distance} $\mathrm{metric}_i$ and a threshold $\tau_i$, both fixed at
commitment time and anchored: a later execution $y_i'$ on the same committed
input, and where applicable the same committed context, reproduces $y_i$ if
and only if $\mathrm{metric}_i(y_i, y_i') \le \tau_i$. This is weaker than the
hash equality our prior system AutonomousSoftwareOrg checks, which rejects
every legitimate re-run of a nondeterministic stage.

\subsection{Adversarial Model}

The node that executed $s_i$, the \emph{recorder}, is not trusted to report
honestly: it may commit an output it did not produce, replay a commitment from
an unrelated execution, or withhold one it dislikes. Signatures are
unforgeable, so a recorder signs only with its own key and an anchored
commitment cannot be altered.

Three properties follow in different strengths, which fixes what the paper
claims. Commitment integrity, that an anchored claim cannot be altered,
backdated, or duplicated under one (recorder, execution identifier) pair,
holds by construction on-chain (Section~5.1); two recorders can still anchor
conflicting claims under one identifier (Section~3.3). Semantic consistency, that a claim falls inside the
committed metric's acceptance region around a re-execution, holds at the rates
Section~\ref{sec:adversarial} measures and yields a local diagnostic rather
than an adjudicated verdict (Section~4.3). Execution provenance, that the
claim is what the stage produced, does not hold at all. Soundness is measured
rather than asymptotic: the error rate is a property of the metric, the
threshold, and the attack. We assume no honest majority;
soundness rests on any party being able to re-execute a committed stage and
challenge it.

\subsection{Partition Model}

A recorder may be disconnected from the ledger for an unbounded but
eventually-ending period, retaining commitments locally and anchoring them
once contact is restored. The delay opens an \emph{equivocation window} in
which it could commit two different outputs for what it later calls one stage
execution, neither yet visible to anyone. A per-recorder monotonic sequence
number, enforced on-chain, prevents replaying or backdating a superseded
commitment and nothing more. Two commitments under validly increasing numbers
can still conflict: neither the sequence number nor a freely chosen identifier
is bound to any execution.

We therefore add two bindings, both enforced on-chain (Section~5.1): each
commitment carries an execution identifier
$\mathrm{eid}_i = H(w \Vert i \Vert H(x_i))$ over the workflow instance $w$,
the stage position $i$, and the committed input, with the registry admitting
at most one commitment per (recorder, $\mathrm{eid}_i$); and each commitment
names its predecessor, so a recorder's commitments form an append-only chain
and anything not extending its head is refused.

The identifier binds a workflow instance and a stage position but not an
execution attempt, so a recorder's second attempt on one input collides with
its first and retries are outside what it can express. Within that limit it
cannot backfill a superseded commitment or anchor two conflicting claims under
one identifier, whatever sequence number the second carries. That is narrower
than one claim per execution. $\mathrm{eid}_i$ is computed offline, and
nothing authenticates $w$ or $i$ as properties of a dispatch, so one run
described under two tuples anchors twice, and two recorders can conflict
about one logical execution. Uniqueness holds only per (recorder,
$\mathrm{eid}_i$); closing either gap needs a dispatcher signature we do not
assume. Two branches
from one predecessor are attributable evidence of a fork once both are
revealed, but a partitioned recorder can sign two and disclose one, leaving
the ledger consistent (Section~\ref{sec:partition}).

\subsection{Trust Assumptions and Non-Goals}

The protocol assumes an eventually reachable ledger, immutable once a
commitment lands, and the unforgeable signatures of Section~3.2. Out of scope: compromise of the model or
tool providers a stage calls into, key management, and collusion among all
challengers, which defeats any optimistic scheme. Two premises are assumed rather than enforced: that the policy was fixed
before the output existed (Section~4.1) and that $\mathrm{eid}_i$ describes the
dispatch it names (Section~3.3). Without them a recorder may pick metric,
threshold, and context after seeing its own output, or describe one run under
two tuples; closing either needs a dispatcher- or requester-signed
precommitment we do not define.

\section{Protocol Design}

\subsection{Commitment at the Stage Boundary}

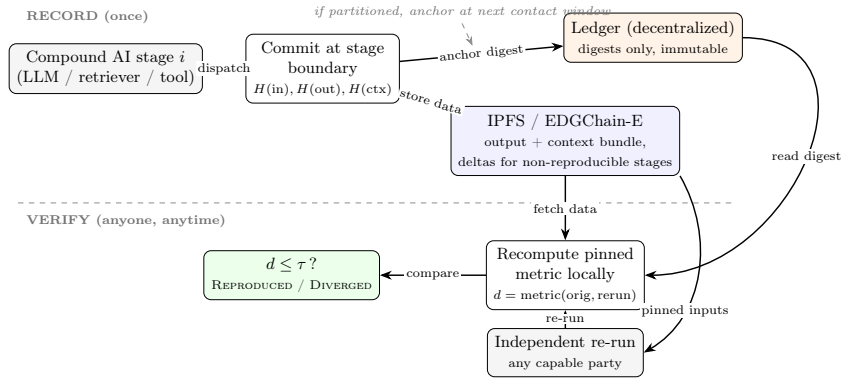
\begin{figure}[htbp]
\centering
\resizebox{0.96\textwidth}{!}{%
\begin{tikzpicture}[
  x=1cm,y=1cm,
  font=\footnotesize,
  box/.style={rounded corners,draw,align=center,inner sep=4pt,minimum height=9mm,minimum width=26mm},
  store/.style={box,fill=blue!6},
  chain/.style={box,fill=orange!10},
  actor/.style={box,fill=black!4},
  verdict/.style={box,fill=green!8},
  >={Stealth[]},
  arr/.style={->,thick},
  lbl/.style={midway,fill=white,inner sep=1pt,font=\scriptsize}
]

\node[actor]  (stage)  at (0,2.0)    {Compound AI stage $i$\\(LLM / retriever / tool)};
\node[box]    (commit) at (4.2,2.0)  {Commit at stage\\boundary\\{\scriptsize $H(\text{in}),H(\text{out}),H(\text{ctx})$}};
\node[chain]  (ledger) at (10.6,2.6) {Ledger (decentralized)\\{\scriptsize digests only, immutable}};
\node[store]  (ipfs)   at (8.9,0.6)  {IPFS\\{\scriptsize output + context bundle,}\\{\scriptsize deltas for non-reproducible stages}};

\node[box]     (recmp) at (8.9,-2.0)  {Recompute pinned\\metric locally\\{\scriptsize $d=\text{metric}(\text{orig},\text{rerun})$}};
\node[actor]   (rerun) at (8.9,-3.5)  {Independent re-run\\{\scriptsize any capable party}};
\node[verdict] (out)   at (3.6,-2.0)  {$d\le\tau$\,?\\{\scriptsize \textsc{Reproduced} / \textsc{Diverged}}};

\draw[dashed,black!45] (-1.7,-0.6) -- (12.2,-0.6);
\node[font=\scriptsize\bfseries,anchor=west,black!55] at (-1.65,3.0)  {RECORD (once)};
\node[font=\scriptsize\bfseries,anchor=west,black!55] at (-1.65,-0.95){VERIFY (anyone, anytime)};

\draw[arr] (stage) -- node[lbl]{dispatch} (commit);
\draw[arr] (commit) -- node[lbl,sloped]{anchor digest} (ledger);
\draw[arr] (commit) -- node[lbl,pos=0.62,sloped]{store data} (ipfs);
\node[align=center,font=\scriptsize\itshape,black!55] (pnote) at (6.9,3.1)
  {if partitioned, anchor at next contact window};
\draw[dashed,black!45,->] (pnote.south) -- (7.1,2.45);

\draw[arr] (ledger.east) to[out=0,in=0,looseness=1.5]
      node[lbl,pos=0.5]{read digest} (recmp.east);
\draw[arr] (ipfs) -- node[lbl]{fetch data} (recmp);
\draw[arr] (ipfs.south east) to[out=-60,in=30]
      node[lbl,pos=0.7]{pinned inputs} (rerun.east);
\draw[arr] (rerun) -- node[lbl]{re-run} (recmp);
\draw[arr] (recmp) -- node[lbl]{compare} (out);

\end{tikzpicture}%
}
\caption{A digest is saved at the stage boundary, before the data path deletes
the intermediate state, and anchored on a decentralized ledger; any party can
later re-run the stage on the committed inputs and recompute the pinned metric
against $\tau$, so the diagnostic needs no trusted arbiter. Content-addressed
storage of the data (shown) is design intent (Section~7).}
\label{fig:protocol}
\end{figure}

The protocol intercepts stage execution at the dispatch boundary and again
when it returns $y_i$ and $c_i$, computing content-addressed digests $H(x_i)$,
$H(y_i)$, $H(c_i)$. These, the recorder's identity, the execution identifier and predecessor of
Section~3.3, and a policy digest
$H(\mathrm{metric}_i \Vert \tau_i \Vert \mathrm{env}_i \Vert \Delta_i)$ form
the commitment. It pins which metric, threshold, environment, and retention
interval apply, over a canonical encoding so two implementations recover the
same digest. A digest authenticates a policy without revealing it, so the
preimage must be served like any other artifact and a challenge is
self-contained relative to the retained set, not the chain. All of this
happens before the data path can discard or relocate $y_i$ or $c_i$. One
commitment is anchored per stage and cannot be complete before the stage
returns. Dispatch fixes $H(x_i)$ and the policy locally, but the chain sees
them already joined to $H(y_i)$. Policy precedence is therefore a property of
the implementation rather than something a verifier checks. A partitioned recorder anchors at the next contact window under the
bindings of Section~3.3, which hold a queued commitment to the order it
declares on submission. Whether that matches the order computed is again
unverifiable: the chain authenticates neither when a stage ran nor when its
commitment was formed.
Figure~\ref{fig:protocol} shows the record and verify lanes end to end.

\subsection{Semantic-Equivalence Commitment}

Nondeterminism is not one phenomenon, so $\mathrm{metric}_i$ and $\tau_i$ are
chosen per stage: sampling variance at nonzero temperature against a semantic
threshold, a pinned-seed stage against a numeric tolerance bounded by
floating-point nondeterminism, a wide-spread stage against a threshold derived
from that spread (Section~\ref{sec:relative}). The hardest case is
world-driven nondeterminism, a live API response or unseeded draw, where no
fixed input determines the output; the protocol records the entropy source
into $c_i$ and injects the realized value in place of the live call, making
the stage deterministic relative to what was recorded. That checks consistency
with the recorded value and never that the external event occurred; where the
value is close to the output, as in our tool stage, little is left to check.

\subsection{The Challenge Procedure}

Verification is optimistic: an anchored commitment stands unless challenged,
and a challenge is a decision procedure with one parameter, the sample size
$k$. The challenger reads the digests and policy from the chain, fetches the
committed artifacts (Section~4.4), and re-executes $s_i$ $k$ times on the
committed $x_i$ and, for world-driven stages, the committed entropy in $c_i$,
accepting if and only if
$\mathrm{median}_{j \le k}\, \mathrm{metric}_i(y_i, y_i'^{(j)}) \le \tau_i$.
A single re-execution cannot distinguish a lying recorder from ordinary
variation, one distant sample rejecting an honest commitment and one close
sample admitting a dishonest one, so the median prices that variance into the
decision instead of letting one draw settle it.

The procedure has no participants: every input comes from the ledger and the
committed artifacts, so any party runs it alone, with no vote or quorum.
There is no shared verdict either. Two challengers drawing fresh samples near
$\tau_i$ can decide differently, and the protocol fixes neither an
authoritative sample set nor a rule for resolving that, nor does a larger $k$
settle it (Section~\ref{sec:adversarial}). No verdict is recorded on chain, and a
challenge is raised only on doubt. No stronger similarity metric closes the
gap: any predicate a challenger can evaluate is one an attacker can aim at, so
a recorder that fabricates inside the metric's acceptance region passes every
challenge. That is semantic consistency and not provenance, and binding a
commitment to an execution rather than an output needs an attestation channel
we do not assume.

\subsection{Artifact Retention}

Anchoring $H(x_i)$, $H(y_i)$, and $H(c_i)$ fixes what was claimed. It does
nothing to keep the claimed bytes reachable. The retained set is four
artifacts: the policy preimage, the input, the output, and the context. Any
one unavailable is as fatal as any other: without the policy a challenger
does not know which predicate to evaluate, without the input it cannot
re-execute, without the rest nothing to compare.
The commitment therefore carries a retention interval $\Delta_i$ in the policy
digest, the deadline being ledger inclusion plus $\Delta_i$. Measuring from
inclusion keeps the rule well defined under deferred anchoring. A commitment
signed before a partition would otherwise arrive already expired, and setting
the deadline at anchor time would change a policy already signed. If
no preimage hashing to a committed value is served within
$\Delta_i$ of inclusion, the diagnostic returns \textsc{Unavailable} with no
re-execution, a third state rather than the divergence a metric comparison
returns (Section~3.1), though the consequence is the same and our
implementation collapses the two onto one verdict string.
It cannot attribute that failure: with no verifiable request time and no
delivery evidence, deliberate refusal, network partition, and the challenger's
own fault are indistinguishable. Availability is an obligation the recorder carries; the protocol assumes
nothing in its place.

\section{Implementation}

The protocol of Section~4 is this paper's contribution; its infrastructure is
reused unmodified from our own prior system (Section~2). The anchor
contract below follows eBlocBroker's coordinator-free scheduling without
extending it.

\subsection{Anchor Contract}

The anchoring mechanism is a minimal Solidity contract,
\texttt{AnchorRegistryV2}, built with Foundry. Each anchored stage stores
eight fields: the digests of $x_i$, $y_i$, and $c_i$, the policy digest, the
predecessor link, the recorder's address, a sequence number, and a timestamp,
keyed by an identifier the recorder computes offline. Three mappings enforce
the three bindings, reverting with \texttt{Sequence\allowbreak NotMonotonic},
\texttt{Execution\allowbreak Already\allowbreak Committed}, or
\texttt{Chain\allowbreak Not\allowbreak Extended}. The execution
identifier and chain link instead ride on the event, recoverable from logs at a
fraction of a stored write's cost. It compiles under Solidity 0.8.35, is
deployed on Sepolia at
\texttt{\seqsplit{0x10F4eB784Ef6441f7Cc3707859a63ceE7DE458b7}}, and costs
216{,}998 gas per anchor; we verified the deployment against all three
bindings, not only the source (Section~\ref{sec:partition}).

\subsection{Synthetic Pipeline}

The pipeline computes the three digests at each stage's boundary and anchors
them. Stage~1 issues a real LLM call through the Claude Code CLI, invoked with
the \texttt{sonnet} alias but committing the identifier the call resolved to,
\texttt{claude-\allowbreak sonnet-\allowbreak 5}, along with the system prompt,
so a challenger reproduces against the model the recorder used whatever the
alias later means. Stage~2 retrieves over HotpotQA's 10 distractor paragraphs
per question using local sentence-transformers embeddings, and Stage~3 draws a
random integer from random.org as committed context. A fixed 50-question
subset anchors every Section~6 experiment, each stage exercising a different
nondeterminism category of Section~4.2: sampling variance, near-tied ranking,
and world-driven entropy.

Pinning the prompt was a repair, not a precaution: an early run let the CLI
inherit the invoking machine's configuration and two honest reproductions of
one question scored 0.95 apart, measuring the host, not the stage's variance. Every semantic result in
Sections~\ref{sec:fidelity}--\ref{sec:partition} postdates the fix and earlier
responses were discarded, not pooled, while Section~\ref{sec:overhead}'s
timings predate it but never compare outputs.
The rest of the environment cannot be reconstructed after commitment either: a
provider may change weights under \texttt{claude-\allowbreak
sonnet-\allowbreak 5}, the CLI inherits machine-level configuration we cannot
fully suppress, and \texttt{all-\allowbreak MiniLM-\allowbreak L6-\allowbreak
v2} is pinned by name alone. The committed $\mathrm{env}_i$ names an
environment without reproducing one: the argument for committing metric,
threshold, and environment together rather than a global constant that
differs by machine, and why a threshold calibrated here needs recalibration
elsewhere.

\subsection{Metric and Artifact Storage}

The equivalence metric is embedding cosine \emph{distance}, $1-\sigma$ for
similarity $\sigma$, under a local offline sentence-transformers model
(\texttt{all-MiniLM-L6-v2}) rather than a proprietary API, so any reviewer
recomputes every Section~6 score without credentials. The harness writes each
stage's retained set to a content-addressed store keyed by the digest the
commitment anchors. The challenge verifies all four are retrievable and
re-hashes what it reads before re-executing, the check a challenger applies to
bytes served by an untrusted recorder. The claimed output used in the
comparison, though, is passed in-process rather than fetched.

\section{Evaluation}
\subsection{Overhead and Baselines}
\label{sec:overhead}

Each stage was run unverified, under the optimistic protocol, and under an
eager variant re-executing immediately. At $N=15$ the anchor alone costs 10.9
to 17.8\,s whether the stage it commits took 12 seconds (the LLM call) or a
fifth of one (the tool call). What it measures is Sepolia block
confirmation, not the protocol: a commitment is complete once the digests
are computed and need not wait on it. Chaining the tool stage to lengths 3, 10
and 30 grew overhead linearly (28.1, 130.0, 473.9\,s), the per-stage drift
from 9.4 to 15.8\,s tracking the nonce queue behind one wallet. Ours anchors
synchronously, hence the linear growth.

\subsection{Semantic-Reproduction Fidelity}
\label{sec:fidelity}

Whether the committed metric separates reproduction from divergence depends on
which data sets the threshold and which measures it. We ran the LLM stage
three times on each of thirty HotpotQA questions, split into two disjoint
halves. Within a half, each question's three answers give all three of its
within-question pairs, 45 across the 15 questions, and each unordered pair of
questions contributes one cross-question pair, the first answer of each,
standing in for a tampered response, $\binom{15}{2} = 105$ in all. Every pair
is weighted equally, so negatives outnumber positives $105{:}45$; pairs never
cross the split, and the threshold is swept on the calibration half and
applied unchanged to the held-out half, fixing $\tau=0.608$ at 98.0\% accuracy
on its own half. On held-out
questions it accepts 44 of 45 honest reproductions and rejects 104 of 105
divergent pairs, 98.7\% accuracy, one error of each kind: false rejection
2.2\%, false acceptance 1.0\%. Those are pair-level rates over pairs that are
not independent, 15 questions at three outputs each, so we count
independent units where the construction supplies them. It does for false
rejection, whose pairs sit inside one question: one of 15 independent
questions erred, 6.7\% [1.2, 29.8] against the [0.4, 11.6] a pair-level
calculation gives. It does not for false acceptance: a cross-question pair spans
two questions and the 105 share endpoints dyadically, so neither the
pair-level [0.2, 5.2] nor a binomial over 15 questions measures the
probability a divergent pair is accepted. We report 1 of 105 and leave the
rate uninterpreted. The split does settle that the populations overlap, intersecting
between 0.577 and 0.651. No threshold separates them, and any $\tau$ trades
one error class against the other.

An earlier version reported perfect separation at an equivalent $\tau$ of
0.359, inflated by sweeping the threshold over the pairs it was scored on and
by an unpinned prompt (Section~5.2), whose pinning moved the held-out
false-rejection rate from 15.6\% to 2.2\%. Two design
limits remain, neither repairable after the fact: $\tau$ and $k$ were not
predeclared, and $\tau=0.608$ maximizes accuracy on a mixture whose
cross-question negatives outnumber same-question pairs more than two to one,
the case any metric separates, so it is optimized against the easy attack.
Section~\ref{sec:relative} asks what a non-constant threshold does instead.

\subsection{Soundness Under an Adversarial Recorder}
\label{sec:adversarial}

Honest and dishonest recorders commit through one funded wallet distinguished
by an off-chain label. What this measures is therefore the challenge
procedure: the identity, sequence, and chain rules were already verified
against the deployed contract (Section~5.1). A dishonest trial commits a claim the recorder did not
produce, internally consistent with its own digest and invisible to hash
comparison, and we vary how hard each tries to look real: a claim
copied off an unrelated question (D1), a confident falsehood about the
question actually asked (D2), and attackers aiming at the decision rule, D3 at
the committed threshold and D4 at the per-execution rule of
Section~\ref{sec:relative}. All three fabricate from one prompt and differ in
what they aim with. D3 re-executes the stage once and submits whichever of 3
candidates sits closest to that probe; D4 re-executes 5 times, computes from
those probes the threshold the challenger will compute, and submits whichever
of 4 candidates has the smallest median distance to them, which maximises its
margin under a rule reading its threshold off the probes alone. Both have the
challenger's own access to the stage and know the rule; neither sees the
re-executions it will draw. Budgets were fixed once and never swept, so these
are two attacker designs and not a boundary over the class.

\begin{table}[htbp]
\caption{Correct verdicts by condition: acceptance for \emph{honest},
rejection otherwise. The $k$ columns apply the committed $\tau=0.608$;
\emph{Per-exec.} applies the rule of
Section~\ref{sec:relative} to the same $k=5$ samples, so the two are compared
on identical data. D4$'$ is D4 restricted to the 10 commitments its probes
did not both predict a loss on and pin the challenger's threshold to within
0.002 of the value realised.
Withholding reports output, input, and policy preimage withheld separately,
5/5 apiece, with no per-execution verdict since \textsc{Unavailable} precedes
re-execution. ``Median dist.'' is the median
across trials of each trial's median re-execution distance at $k=5$.}
\label{tab:adversarial}
\centering
\footnotesize
\begin{tabular}{lccccc}
\hline
\textbf{Condition} & \textbf{$k=1$} & \textbf{$k=3$} & \textbf{$k=5$} & \textbf{Per-exec.} & \textbf{Median dist.} \\
\hline
Honest                       & 15/15 & 15/15 & 15/15 & 15/15 & 0.088 \\
D1 cross-question            & 14/14 & 14/14 & 14/14 & 13/14 & 0.967 \\
D2 same-input fabrication    & 3/29  & 3/29  & 2/29  & 19/29 & 0.240 \\
D3 threshold-aware adaptive  & 3/15  & 2/15  & 2/15  & 8/15  & 0.181 \\
D4 rule-aware adaptive  & 0/15  & 0/15  & 0/15  & 11/15 & 0.255 \\
D4$'$ selective (derived)    & 0/10  & 0/10  & 0/10  & 6/10  & 0.227 \\
W withholding (each target)  & 5/5   & 5/5   & 5/5   & ---   & --- \\
\hline
\end{tabular}
\end{table}

What separates the conditions is the attack, not $k$. Every honest commitment
was accepted (100\% [80, 100]) and every cross-question substitution rejected
at all three sample sizes (100\% [78, 100]), median distance 0.967 against a
threshold of 0.608; earlier versions tested that condition alone, which would
have supported a 100\% detection claim.

Same-input fabrication is another matter: asked to answer the question
actually posed, confidently and wrongly, the recorder got away with it in 26
of 29 trials at $k=1$ and 27 of 29 at $k=5$, a detection rate of 10\% [4, 26]
falling to 7\% [2, 22]. One commitment inverted the order and dates of two
Civil War battles against a re-execution that got them right, scoring 0.115
apart: an embedding distance measures whether two texts are about the same
thing, and a factual inversion is about the same thing. The threshold-aware attacker fared about the same at one sample, 20\% [7,
45], and better as $k$ rose, 13\% [4, 38]. These intervals need no clustering
adjustment, since each trial uses a distinct question; the dependence left is
across conditions, D2's 29 questions containing all 15 of D3's and 14 of D4's,
D1's 14 being disjoint.

Larger $k$ did not help: the adaptive attacker's 3 of 15 falling to 2 of 15 by
one trial sits within sampling variation, not evidence that samples hurt.
Repeated sampling reduces the variance of the estimated distance without
moving it, so a fabrication whose population median already sits below $\tau$
is accepted more confidently. That bounds the claim: no $k$ repairs a fixed
threshold on the accepted side of it, though a fabrication whose median sits
above $\tau$, a case none of our conditions produced, is one more samples would
reject. A different constant helps but calibration cannot
reach it: Section~\ref{sec:fidelity}'s sweep ranks 0.608 one pair in 150 above
the tighter point Section~\ref{sec:relative} matches against. That point has
priced the easy attack, and choosing it needs the attack outcomes a
deployment does not have. The procedure decides semantic divergence, and
same-input fabrication is not that.

Withholding is the one dishonest strategy the diagnostic decides outright. In
5 trials each for a withheld output, input, and policy preimage, a recorder
that anchored a real commitment and then refused to serve one artifact was
judged \textsc{Unavailable} at once; these rows check the decision rule, not a
detection rate.

\subsection{A Per-Execution Threshold}
\label{sec:relative}

The failures above share a cause the metric does not account for on its own.
A single $\tau$ must be loose enough for the noisiest execution the protocol
will judge. That makes it far too loose for any particular one: honest
re-executions sit a median 0.088 apart, and $\tau=0.608$ admits everything out
to seven times that. We therefore evaluated a second rule on the same
committed data, deriving the threshold from the stage's own variance:
accept only if the claim's median distance to the $k$ re-executions falls
within one standard deviation of how far they sit from each other,
\[
  \mathrm{median}_{j \le k}\; \mathrm{metric}_i\!\left(y_i, y_i'^{(j)}\right)
  \;\le\; \mu(D) + \kappa\,\sigma(D),
\]
where $D$ collects the $\binom{k}{2}$ pairwise distances
$\mathrm{metric}_i(y_i'^{(a)}, y_i'^{(b)})$, $\mu$ and $\sigma$ are their mean
and population standard deviation, $\kappa = 1$, and $k \ge 3$. The non-strict
inequality of Section~3.1 makes a zero-spread stage an exact match. Nothing
here is calibrated against a mixture of attacks. $\kappa$ was fixed at one
before any trial here, on an earlier sample this evaluation
replaced, so the D2 column is held out rather than fit.

The rule detects same-input fabrication in 19 of 29 trials, 66\% [47, 80],
against 2 of 29 for the committed threshold on the same $k=5$ samples,
rejecting no honest commitment, 15 of 15, and costing one D1 trial, 13 of 14.

A tighter rule detects more, so the difference has to be read at a matched
operating point. \texttt{analysis/matched\_operating\_point.py} sweeps the
fixed threshold over every point it can occupy on these trials and reads it
where it, too, rejects no honest commitment: $\tau = 0.437$, reaching 9 of 29
on D2, 3 of 15 on D3, and 2 of 15 on D4, against 19, 8, and 11; allowed 2
false rejections in 15, it reaches 13 of 29. The comparison is generous to
the constant, chosen with the honest and the attack outcomes both in hand and
scored on the same 15 honest trials, while $\kappa$ was fixed before these
trials existed.
D2 and D3 predate the rule, so that column is recomputed from the stored
re-execution texts by \texttt{analysis/relative\_rule.py}, which agrees with
the stored verdict in every D4 trial when run with the rule live.

The two rules are not independent samples, both read off the same five
re-executions, so \texttt{analysis/paired\_test.py} tests the discordant
trials with McNemar's exact test rather than comparing intervals: significant
on D2 (19 moved from missed to detected against 2 the other way,
$p=2.2\times10^{-4}$) and D4 (11 against none, $p=9.8\times10^{-4}$), not on
D3 (7 against 1, $p=0.070$).

Both sides of this predicate move with $k$, so we swept it over the same
stored samples, taking the first $k$ re-executions. Detection barely
moves: 19, 20, 19 of 29 on D2 at $k=3,4,5$; 8, 9, 8 of 15 on D3; 10, 12, 11 of
15 on D4; and 13 of 14 on D1 throughout. The false-rejection rate does not.
Honest commitments were accepted 12 of 15 at $k=3$ and 13 of
15 at $k=4$, against 15 of 15 at $k=5$. The threshold is estimated from
$\binom{k}{2}$ pairwise distances, 3 of them at $k=3$ against 10 at
$k=5$, so a sparse estimate of the spread comes out too tight often enough to
accuse an honest recorder. This inverts what $k$ did for the fixed threshold,
where more samples bought nothing. The sweep is nested rather than
independent: the samples at $k=3$ are a prefix of those at $k=5$, and the
$\binom{k}{2}$ distances share re-executions instead of supplying fresh ones.
So three changed outcomes over 15 honest trials show a rule sensitive to $k$,
not a false-rejection rate it controls. What they do fix is that the rule's
clean honest column belongs to $k=5$, and that a policy pinning $\kappa$ must
pin $k$ beside it.

A public rule can be aimed at as readily as a public constant, and D4
(Section~\ref{sec:adversarial}) is built to do exactly that. It carries one
handicap it cannot remove: the challenger's threshold comes from
re-executions drawn after the commitment exists, so D4 estimates, from an
independent sample of its own, a target not yet realized. It was detected in
11 of 15 trials, 73\% [48, 89], while the same 15 commitments were detected 0
of 15 by the committed threshold, [0, 20]: what a fixed constant is worth
against an adversary who knows it. D2's 66\% against this one's 73\%
measures nothing: the two attacks fabricate differently, over question sets
that only partly overlap.

Two properties of the rule, not of the attacker, carry this. It is tighter:
in 8 of the 11 detections the attacker's own estimate of its distance already
exceeded the threshold the challenger went on to compute, so exact knowledge
of that threshold would not have rescued its claim. The other 3 turn on a
target not yet realized. Paired against the value each trial went on to
produce, the attacker's threshold estimates carry no systematic error we
can see: median difference 0.000 over 15 trials, with 7 above, 5 below, and 3
exact. But they are wrong trial by trial, mean $+0.027$ and standard
deviation 0.072 against thresholds whose own median is 0.184. It expected to
clear in 7 of the 15 trials; 3 of those 7 did. A fourth trial cleared too,
one it had expected to lose.
Tightness is sharpest where the stage was near-deterministic for that input,
collapsing the threshold toward zero: 5 trials, all 5 caught. Our attacker
submits on every question, but need not. On those five, its probes pinned the
threshold within 0.002 of the value realised and predicted failure, so one
free to abstain skips them at no cost, giving the D4$'$ row of
Table~\ref{tab:adversarial}, 6 of 10 rather than 11 of 15. Its probes
predicted a loss on 3 further trials without pinning the threshold that
closely; abstaining on those as well leaves 4 of 7, a denominator too small
to read.

Three things this does not establish. The rule was evaluated on the data these
conditions produced, not committed in any anchored policy digest: every
commitment in Table~\ref{tab:adversarial} pins $\tau=0.608$, and a protocol
adopting the rule would have to pin it entire, $\kappa$ with $k$, the
construction of $D$, the dispersion convention, and the comparison operator.
Nor is it calibrated: $\kappa=1$ carries no false-rejection guarantee, and the
sweep above shows what that costs once $k$ falls. And two attacker designs do
not exhaust the class. What it establishes is narrower. On identical samples,
holding the cosine metric and this pipeline's re-execution spread fixed,
changing only what the claim is compared against moved detection where no $k$
did, so the boundary of Section~\ref{sec:adversarial} is set by the fixed
threshold, not the sample size. Another metric, or a stage with a different
re-execution spread, relocates it, and both rules still accept fabrications,
the 0.115 factual inversion among them.

\subsection{Partition Tolerance and Fork Behaviour}
\label{sec:partition}

A real LEO contact gap runs to tens of minutes, so a five-second sleep stands
in: under test is what a recorder can and cannot do across a gap, not its
duration. Four cases ran against the deployed registry. Deferral works: three
commitments computed while disconnected were held locally, then anchored on
reconnect under sequence numbers 86, 87, 88, order preserved, at delays of
35.9, 36.4 and 37.0 seconds, again Sepolia confirmation latency. Equivocation
is refused: a conflicting claim about the \emph{same} execution was rejected
with \texttt{Execution\allowbreak Already\allowbreak Committed} even carrying
a strictly larger sequence number, the case Section~3.3 calls unbounded under
sequence numbers alone. A reordered submission failed too, with
\texttt{Chain\allowbreak Not\allowbreak Extended}. The fourth is a negative
result: a partitioned recorder signed two branches against one predecessor and
disclosed one, the chain staying consistent, no anchor-time rule detecting it
since the evidence of a fork is the branch that never arrives.

Together the four establish something narrower than partition tolerance,
ordered immutability conditional on eventual submission and the survival of
queued local state, not liveness against the recorder itself: retention runs
from ledger inclusion (Section~4.4), so an unfavourable commitment can be
dropped without starting a clock or leaving a trace. A stronger claim needs a
bounded anchoring obligation and recovery tests over lost queued state.

\section{Discussion and Limitations}

This is a proof-of-concept, not a deployment study, and three limitations
follow. First, the pipeline is synthetic: a three-stage HotpotQA sequence in
one process, not across live edge, cloud, and orbital nodes, so the partition
cases exercise a deferred anchor and not distributed retrieval, a challenge
run across a gap, or recovery of queued state after a restart. Connected-state
latency comes from HyperDrive's published figures, but neither publishes
contact-window durations, so those come from LEO orbital-mechanics
literature rather than a live trace. The ledger is
a public Sepolia testnet; a permissioned chain would trade censorship
resistance for lower latency.

Second, the metrics cover this pipeline's three nondeterminism categories and
not every category a stage could exhibit, soundness holding relative to the
committed metric rather than to correctness. Two mechanisms are verified but
not stress-tested. The execution identifier and predecessor chain are
exercised against one recorder on one testnet and serialize by construction,
so a stalled predecessor holds up what queues behind it. The retention rule
is backed by a local store that assumes away the distributed-availability
problem IPFS faces.

Third, the protocol establishes plausibility under a committed metric, not
execution provenance, which needs an attestation channel. Making the
diagnostic adjudicated needs machinery this paper does not provide either: the
pinned rule and challenge randomness of Section~\ref{sec:relative}, a verdict
state on chain, a verifiable request time and delivery evidence so withholding
can be attributed rather than observed, and the dispatcher signature of
Section~3.3.

Two gaps admit concrete answers. Recast as a conformal $p$-value in an
anytime-valid $e$-process, Section~\ref{sec:relative}'s hand-fixed $\kappa$
would bound the false-accusation rate and let a challenger stop once the
evidence allows. The fork of Section~\ref{sec:partition} cannot be caught at
commit time, but witnesses countersigning and gossiping the recorder's claimed
head would detect a withheld branch once one witness is honest and reachable.

\section{Conclusion and Future Work}

This paper presented a protocol addressing all three conditions prior work
covers at most two of at once (Section~1): each stage's inputs, outputs,
context, and policy committed at its boundary, anchored under an execution
identifier and predecessor chain, and challenged on the median of $k$
re-executions without a quorum. Its most useful finding is the boundary. Every
cross-question substitution and every withholding attempt was caught, while a
fabrication close under the committed metric got through in most trials, and
no sample size repairs that. The threshold does: on identical $k=5$ samples,
holding the metric and this pipeline's re-execution spread fixed, a
per-execution threshold moves detection from 2 of 29 to 19 of 29 without
rejecting an honest commitment, a freedom that is $k=5$'s and not the rule's,
where the best constant matched to the same zero rejections reaches 9 with the
attack outcomes already in hand. It catches 11 of 15 of an attacker built
against it, 6 of 10 once that attacker may abstain, where the committed
constant catches none and the matched one 2. That gap does not close by
sharpening the metric: any predicate a challenger can evaluate is one an
attacker can aim at. Beyond Section~7's two extensions, two directions remain:
live HyperDrive and Currus deployments, and a real contact-window trace in
place of literature-derived parameters.

\begin{credits}
\subsubsection{Availability.}
Contract, pipeline, and the analysis scripts behind every number in Section~6:
\url{https://github.com/avatar-lavventura/acism-verification}.
\subsubsection{\discintname}
The author has no competing interests to declare that are relevant to the content of this article.
\end{credits}

\bibliographystyle{splncs04unsrt}
\bibliography{references}

\end{document}